\documentclass[manuscript,authorversion,nonacm]{acmart} 
\AtBeginDocument{%
  \providecommand\BibTeX{{%
    \normalfont B\kern-0.5em{\scshape i\kern-0.25em b}\kern-0.8em\TeX}}}

\usepackage{graphicx}
\usepackage[caption=false]{subfig}
\usepackage[font=small,labelfont=bf]{caption}
\usepackage{hyperref}
\usepackage{enumitem}
\usepackage{amsthm}
\usepackage{multirow}

\setlist{nolistsep,leftmargin=*}
\begin{document}

\title[Student-AI Collaborative Hint Writing]{Bridging Learnersourcing and AI: Exploring the Dynamics of Student-AI Collaborative Feedback Generation}


\author{Anjali Singh}
\affiliation{%
  \institution{University of Michigan}
  \country{USA}}
\email{singhanj@umich.edu}

\author{Christopher Brooks}
\affiliation{%
  \institution{University of Michigan}
  \country{USA}}
\email{email}

\author{Xu Wang}
\affiliation{%
  \institution{University of Michigan}
  \country{USA}}
\email{email}

\author{Warren Li}
\affiliation{%
  \institution{University of Michigan}
  \country{USA}}
\email{email}

\author{Juho Kim}
\affiliation{%
  \institution{KAIST}
  \country{South Korea}}
\email{email}

\author{Deepti Wilson}
\affiliation{%
  \institution{University of Michigan}
  \country{USA}}
\email{email}








\begin{abstract}

\end{abstract}







\begin{abstract}
This paper explores the space of optimizing feedback mechanisms in complex domains such as data science, by combining two prevailing approaches: Artificial Intelligence (AI) and learnersourcing. Towards addressing the challenges posed by each approach, this work compares traditional learnersourcing with an AI-supported approach. We report on the results of a randomized controlled experiment conducted with 72 Master’s level students in a data visualization course, comparing two conditions: students writing hints independently versus revising hints generated by GPT-4.
The study aimed to evaluate the quality of learnersourced hints, examine the impact of student performance on hint quality, gauge learner preference for writing hints with or without AI support, and explore the potential of the student-AI collaborative exercise in fostering critical thinking about LLMs.
Based on our findings, we provide insights for designing learnersourcing activities leveraging AI support and optimizing students' learning as they interact with LLMs.
\end{abstract}

\maketitle

\section{Introduction}
Timely and effective feedback is essential for supporting learning \cite{shute2008focus}. 
To overcome the challenges in providing personalized feedback at scale many tools rely upon Artificial Intelligence (AI) \cite{balse2023investigating, pankiewicz2023large}. The advent of conversational Large Language Models (LLMs) such as GPT-4 \cite{brown2020language} has fueled new interest in using AI directly for feedback. However, recent work has shown that providing LLM-generated feedback to learners can make them over-reliant on such support \cite{pankiewicz2023large}. Moreover, LLMs suffer from the `black box' phenomenon, leading to a lack of transparency and limited interpretability of their outputs \cite{yan2023practical}. Limitations of these models to answer domain-specific prompts truthfully \cite{balse2023investigating} have necessitated expert oversight for incorporation into educational settings.

An alternative approach for scaling educational feedback is to leverage the power of the learners themselves, via a method referred to as learnersourcing \cite{kim2015learnersourcing}. In this approach learners are engaged in the generation of educational content or assessments, furthering their own understanding of the course content while supporting the learning of their peers. 
Learnersourcing can be impactful not only for generating learning resources at scale, but also for the learning benefits it provides to the students who contribute learning resources, as doing so requires activation of higher-order thinking skills \cite{khosravi2021charting}. However, complex learnersourcing tasks can be challenging for students, especially when they lack task-specific knowledge \cite{singh2022learnersourcing}. Consider the learnersourcing task of writing a hint as a form of formative feedback to a peers' incorrect assignment submission: first the learner must identify the set of mistakes in the incorrect solution and map those mistakes to misconcenceptions. Then they must determine appropriate domain-based remediation strategies.
Since providing this strategy directly undermines the hint receiver's knowledge construction, the learner writing the hint must provide feedback to address the mistakes without giving away the full solution. Finally, considerations related to the delivery of the hint (e.g., tone and specificity) must be considered. In short: writing good hints is difficult, which can lead to the generation of poor-quality learnersourced feedback. 


In this work, we explore the dynamics of student-AI collaborative hint writing for generating high quality hints while providing student hint-writers with a meaningful learning experience. While hints generated by LLMs, when properly prompted, can be useful for providing task-specific knowledge, it is unclear how the accuracy and specificity of LLM-generated hints impact the quality of learnersourced hints. For instance, LLMs often provide verbose rather than specific responses, which can be distracting to learners.
To this end, this work compares the learnersourcing task of writing hints when learners are given LLM-generated hints to improve versus when they write hints on their own.

We situate our research in the domain of data science education with the goal of generating high-quality formative feedback on students' incorrect programming assignment solutions. 
We conducted a randomized controlled experiment with 72 students in a Master's level data visualization course. The core learning activity in this experiment engaged students in writing a hint for an incorrect solution to a programming assignment that they had recently worked on, by comparing it to a correct solution. We had two experimental conditions: (1) \textit{Hint-writing}, where students wrote a hint for the incorrect submission by themselves, and (2) \textit{AI-hint-revision}, where students wrote a hint after being shown one generated by the LLM GPT-4 \cite{brown2020language}. While we engaged in prompt engineering with GPT-4 for this task, we are not aiming to answer the question of whether LLM-generated hints can be used directly per se, as the value of learnersourcing goes beyond feedback generation and provides metacognitive benefits to hint writers. Instead, we are interested in the interaction between the learners engaged in learnersourcing and the LLM-generated hint, and thus  
investigated the following research questions:
\begin{itemize}
\item \textbf{RQ1.} What is the difference between the quality of hints when learners write them independently versus when they revise a hint written by GPT-4?

\item \textbf{RQ2.} How is the quality of learnersourced hints impacted by students’ course performance?

\item \textbf{RQ3.} Which design (with or without AI support) do the students prefer? 

\item \textbf{RQ4.} To what extent do the students find the \textit{AI-hint-revision} variant of the exercise helpful for thinking critically about the accuracy and appropriateness of responses provided by GPT-4?
\end{itemize} 

This work contributes to our understanding of how student interactions with AI-based tools, specifically LLMs, impact the outputs of learnersourcing tasks. As students, teachers, and technology platforms rapidly adopt LLM technologies, it is important to understand how this adoption modifies the artifacts of the learning environment (RQ1), whether student performance is a factor in any changes observed (RQ2), whether student motivational preferences, and thus their willingness to engage in future learnersourcing tasks, are affected by such changes (RQ3), and how the AI supported task itself modifies student perceptions of learning in learnersourcing contexts (RQ4).
\vspace{-2mm}

\section{Related Work}

\subsection{Reflecting on Mistakes as a Learning Activity}
Reflecting on mistakes plays a pivotal role in enhancing memory, attention, and active learning. \cite{metcalfe2017learning}. Prior research has highlighted the learning benefits from contrasting incorrect methods with correct ones, such as focusing students' attention on the distinguishing features of the correct examples and the underlying concepts \cite{durkin2012effectiveness}, and improving procedural flexibility \cite{rittle2011power}. In addition to identifying mistakes, the task of hint-writing involves \textit{explaining} the mistakes and how to fix them, without giving away the full solution. 
Prompting students to explain learning materials not only enhances their comprehension by promoting deep-level cognitive processes (e.g., organization and integration of information \cite{fiorella2016eight}), but can also help elicit metacognitive processes which are conducive to enacting effective cognitive strategies \cite{fukaya2013explanation}. Elaborating on mistakes in domains prioritizing problem-solving has been found to substantially augment students' learning outcomes
\cite{loibl2019make}. Data visualization, and in general data science, is one such domain where students concurrently learn effective ways of problem-solving and communicating data-driven insights. Actively engaging learners in writing constructive hints for incorrect solutions to problems they recently solved can encourage them to think deeply about  data science code and compare different approaches to solving the same problem.


\subsection{Learnersourcing and AI for Generating Educational Content}
Learnersourcing, as defined by Kim \cite{kim2015learnersourcing}, refers to a pedagogically supported form of crowdsourcing in which ``learners collectively generate useful content for future learners while engaging in a meaningful learning experience themselves". Learnersourcing has been shown to positively impact learning for both the learners who learn from the learnersourced artifacts, and the learners who create them \cite{singh2022learnersourcing}. For generating hints and explanations at scale, learnersourcing has been leveraged in several ways. Guo et al. prompted learners write an explanation right after they overcame a programming struggle \cite{guo2020learnersourcing}. Glassman et al. \cite{glassman2016learnersourcing} created workflows involving learners reflecting on an obstacle that they recently overcame and comparing their solution to those of other students (with both better and worse solutions) to generate a hint for improving the relatively worse solution. Despite its many benefits, learnersourcing is impeded by challenges such as lack of motivation among contributors which can lead to the generation of low quality artifacts  \cite{singh2021s}. 

Concurrently, researchers exploring the use of LLMs for generating feedback have found that LLMs like GPT-3.5 are able to generate helpful personalized hints for students solving programming assignments, but cautioned that students may over-rely on such feedback \cite{pankiewicz2023large}. Others have reported issues such as high variability in the accuracy of identified mistakes and suggested fixes in the generated feedback \cite{balse2023investigating}. One potential solution is to leverage human-AI partnerships involving humans-in-the-loop to both provide inputs to and evaluate and improve the LLM generated artefacts \cite{denny2022robosourcing}. We explore this opportunity by engaging learners in a data visualization course in evaluating and revising LLM generated hints for incorrect programming assignment solutions.

\section{Method}

\subsection{Course Context and Study Population}
The study was conducted in an online Data Visualization course offered to graduate students as part of an online Masters of Applied Data Science program by the University of Michigan. This program consists of adult learners (mean age = 32.50 years, SD = 7.99), many of whom are mid-career professionals looking to acquire data science skills. Students in this program are required to have introductory programming and statistics knowledge. The JupyterLab computational notebook environment ~\cite{jupyter} and Python are used for the programming component of this course.
The course had four weekly programming assignments that were manually graded by the instructional team. 

\subsection{Experiment Design}
We conducted a mixed-model design experiment (consisting of both a within-subjects and a between-subjects component) with a crossover design. 
All students were required to engage in two `reflections' that instructed them to write a hint for an incorrect solution to a programming assignment that they had recently completed. For reflection-1 (given in the first week of the course), approximately half the students were instructed to revise a GPT-4 generated hint for the incorrect solution (\textit{AI-hint-revision} condition), while the others were instructed to write a hint independently (\textit{Hint-writing} condition). In reflection-2 (given in the second week), the roles reversed: those who previously didn't receive the GPT-4 hint now did, and vice versa. This experiment design was chosen to gauge the difference between the quality of hints generated by students in the \textit{AI-hint-revision} versus \textit{Hint-writing} conditions, as well as understand each student's preference of writing hints with versus without AI support.


\subsection{Reflection Assignments}
Each reflection was graded (worth 2.5\% of the total course grade), and was based on the programming assignment for that week.
Students received full points for completing the reflection, and no points otherwise. In each reflection, we first described the task to students: identifying the mistakes in an incorrect solution to a question from the most recent programming assignment by comparing it against a correct solution, and writing a hint such that someone who wrote the incorrect solution could use it to identify their mistakes and fix their code. For each reflection, the same incorrect and correct solutions were shown to all students. Students in the \textit{AI-hint-revision} condition were informed that they would be shown a hint generated by GPT-4 to help them in this task, and it would be their job to validate if it is correct and appropriately phrased. Next, students in both conditions were informed of the potential learning benefits of this exercise, i.e., encouraging them to think critically, learning from mistakes, and improving their problem-solving skills. Importantly, the incorrect solution for reflection-1 had one major mistake (e.g., incorrectly defining the mean and standard deviation when plotting the t-distribution)  and one minor mistake (e.g., not labeling the axes), while for reflection-2 the incorrect solution had one major and two minor mistakes\footnote{The instructions for each reflection can be viewed \href{https://osf.io/vgrfk/?view_only=8579adb171a64c07bb6641be67dba202}{here}}. 


\subsubsection*{Design of Hint-Writing Task}
As it is recommended to provide students with worked examples of complex learnersourcing tasks involving multiple skills \cite{doroudi2016toward}, 
we provided a simple example based on labeling plot axes to demonstrate what they were expected to do in the hint-writing exercise. 
After going through the worked example, students proceeded to the main exercise. The programming assignment instructions were reproduced and they were shown a correct and an incorrect solution (which included both the code and the resulting visualization plot) for that assignment side-by-side. After this, students in the \textit{Hint-writing} condition were asked to write a hint for the incorrect solution, and students in the \textit{AI-hint-revision} condition were shown the GPT-4 hint and asked to verify and rewrite the hint as required. Learners who were shown the GPT-4 hint were also warned that it could be incorrect, incomplete, or both.

\subsection{Prompting GPT-4 for Hints}
To generate the LLM hints for each reflection, we prompted GPT-4\footnote{The full prompt text is available \href{https://osf.io/vgrfk/?view_only=8579adb171a64c07bb6641be67dba202}{here}} with almost identical instructions as those provided to the students to write their hint, including the worked example. 
After some experimentation with prompting (for instance, asking GPT-4 to write concise hints when it produced verbose hints and not referring to the correct solution provided in the prompt in the hint text) we selected the first hint that GPT-4 generated to our final prompt. 

\subsection{Data Collection and Outcome Measures}
\label{measures}
The reflections were created and distributed using the Qualtrics\footnote{https://www.qualtrics.com/} platform. To evaluate the hints written by learners we qualitatively coded each hint using the criteria described below. These criteria are based on prior literature on desirable qualities of formative feedback \cite{shute2008focus, thurlings2013understanding}. For each criterion, we assigned each hint a binary value 0 or 1 (with 1 being better than 0), except for the metric `\textit{Utility}' for which we rated each hint on a 1--7 ordinal scale. The incorrect solutions for each assignment consisted of one major mistake, which we refer to as the `primary mistake'. Accordingly, the first author defined the criteria for evaluating each hint with the help of another author. Both authors, who had prior experience in assessing data science learners, held multiple rounds of discussions to define each criterion and finalize the codebook. Each author independently coded a random sample of ~35\% of all generated hints using the codebook, discussed cases of disagreement, and iterated thrice on the rubric until strong agreement was achieved (>= 0.8 Cohen's Kappa) for each criterion. Before the final round of coding, we removed any extra text from hints that students had included, such as "Hint:" or references to the GPT-4 hint that could be easily separated from the main hint text (e.g., ``The ChatGPT hint is adequate"). We removed the data for three students who did not write any meaningful hints and wrote text that was irrelevant to the course material (e.g., ".."). Finally, the first author coded all the hints based on the following criteria: (1) \textit{Full Accuracy}: Whether \textit{all mistakes} were identified or not, (2) \textit{Partial Accuracy}: Whether \textit{at least one mistake} was identified or not, (3) \textit{Extraneous Information}: Whether the hint contained any information that was \textit{not} necessary for fixing the mistakes, (4) \textit{Specificity}: Whether the hint for the primary mistake was supported with details without giving away the full solution and explained how/where the incorrect solution did not meet the assignment goals, (5) \textit{Phrasing}: Whether the hint is pedagogically appropriately phrased, using a positive and encouraging tone rather than a harsh or critical tone, and (6) \textit{Utility},
which had the following levels:
\begin{enumerate}
    \item The hint does not cover the primary mistake and requires modification to make it more specific and/or pedagogically appropriate by improving its phrasing and/or removing any extraneous information.
    \item The hint does not cover the primary mistake. Otherwise it is specific and well phrased.
    \item The hint covers at least the primary mistake. It requires modification to make it both more specific, and more pedagogically appropriate by improving its phrasing and/or removing any extraneous information.
    \item The hint covers at least the primary mistake. It requires modification to make it either more specific or improve its phrasing or remove any extraneous information.
    \item The hint covers at least the primary mistake. It is specific, well phrased and has no extraneous information.
    \item The hint is fully accurate, but requires modification to make it either more specific or improve its phrasing. Any extraneous information is not detrimental to students' learning.
    \item The hint is fully accurate, specific, appropriately phrased and does not have any extraneous information
\end{enumerate}
\noindent Additionally, we captured the time spent on writing each hint.
\subsection{Understanding Students' Learning Experience}
We collected students' responses to three survey questions to learn about their experience of working on the reflections. After each reflection, we asked students if they would be interested in engaging in similar exercises in future courses, with `Yes/No' as the response options. We asked students in the \textit{AI-hint-revision} condition to indicate the extent to which the exercise helped them think critically about the accuracy and appropriateness of responses generated by GPT-4 on a 5-point Likert scale. Finally, after reflection-2, we asked them which design (with or with the GPT-4 scaffold) they preferred. They were also given the option to explain the reasons behind their preference.

\subsection{Data Analysis}
To understand the effect of the experimental condition and students' course performance on the hint quality and time spent, we used: (1) a Generalized Linear Mixed Effects model for the outcome variables with binary values, (2) a Linear Mixed Effects model for time spent, and (3) a Cumulative Link Mixed Model for \textit{Utility} which had ordinal values in the range 1--7. We measured students' course performance based on their scores in the manually graded programming assignment-2, which reflected their performance in the course at the time the study was being conducted. Specifically, we looked at the quartiles in which each students' assignment-2 scores belonged. In each of the models, the experimental condition, course performance, and the reflection number (1 or 2) were used as the fixed effects and the student IDs were used as the random effects to take into account the variability amongst students. We also included an interaction term for experimental condition and course performance to investigate the differential effect of condition on students at different levels of course performance. All of these analyses were conducted in R using the \texttt{lme4} \cite{bates2009package} and \texttt{ordinal} packages \cite{christensen2018cumulative}. To analyze the responses to the survey questions regarding students' learning experience and preference for task design, we only included the data of those students who completed both reflections. To understand the reasons behind students' preference for task design, the first author qualitatively analyzed their open ended responses, categorized them into themes and refined the themes after multiple readings of the data. 

\section{Results}
 
\subsection{Data Overview}

Out of a total of 78 students in the course, 63 and 60 students submitted a meaningful response to reflection-1 and reflection-2 respectively. 51 students completed both reflections and 72 students completed at least one reflection.  Table \ref{tab:desc} shows the number of students per condition along with the means and standard deviations for the outcome variables \textit{Full Accuracy}, \textit{Extraneous Information}, \textit{Phrasing}, \textit{Specificity}, \textit{Utility}, and time-spent for each reflection. 

\begin{table}[]
\centering
\footnotesize
\begin{tabular}{lllllllll}
\hline
Reflection & Condition & \# Students & \begin{tabular}[c]{@{}l@{}}Full Accuracy\\ (0-1)\end{tabular} & \begin{tabular}[c]{@{}l@{}}Extraneous\\ Info (0-1)\end{tabular} & \begin{tabular}[c]{@{}l@{}}Phrasing\\ (0-1)\end{tabular} & \begin{tabular}[c]{@{}l@{}}Specificty\\ (0-1)\end{tabular} & \begin{tabular}[c]{@{}l@{}}Utility\\ (1-7)\end{tabular} & \begin{tabular}[c]{@{}l@{}}Time-spent \\ (seconds)\end{tabular} \\ \hline
\multicolumn{1}{l|}{\multirow{3}{*}{Reflection-1}} & \multicolumn{1}{l|}{Hint-writing} & 31 & 0.48 (0.5) & 0.77 (0.43) & 0.77 (0.43) & 0.52 (0.51) & 4.90 (1.68) & 287.82 (413.82) \\ \cline{2-9} 
\multicolumn{1}{l|}{} & \multicolumn{1}{l|}{AI-hint-revision} & 32 & 0.66 (0.48) & 0.84 (0.37) & 0.75 (0.44) & 0.59 (0.50) & 5.41 (1.85) & 335.48 (319.37) \\ \cline{2-9} 
\multicolumn{1}{l|}{} & \multicolumn{1}{l|}{GPT-4 hint} &  & 1 & 0 & 1 & 1 & 6 &  \\ \hline
\multicolumn{1}{l|}{\multirow{3}{*}{Reflection-2}} & \multicolumn{1}{l|}{Hint-writing} & 36 & 0.19 (0.40) & 0.67 (0.48) & 0.61 (0.49) & 0.33 (0.48) & 3.72 (1.50) & 407.43 (538.93) \\ \cline{2-9} 
\multicolumn{1}{l|}{} & \multicolumn{1}{l|}{AI-hint-revision} & 24 & 0.25 (0.44) & 0.46 (0.51) & 0.75 (0.44) & 0.59 (0.50) & 4.00 (1.44) & 515.70 (591.88) \\ \cline{2-9} 
\multicolumn{1}{l|}{} & \multicolumn{1}{l|}{GPT-4 hint} &  & 0 & 0 & 1 & 1 & 4 &  \\ \hline
\end{tabular}
\caption{Number of students per condition and mean (standard deviation) values for Full Accuracy, Extraneous Information, Phrasing, Specificity (each coded on ordinal 0-1 scale), Utility (coded on ordinal 1-7 scale) and Time-spent (in seconds) for each reflection}
\vspace{-5mm}
\label{tab:desc}
\end{table}

There was no significant difference based on prior knowledge of students between conditions as measured by their assignment-1 scores (p=0.30). 
Additionally, for each reflection, we calculated the normalized counts (\textit{NC}) of students per each tercile\footnote{We used terciles as the Assignment-1 scores could not be distributed into quartiles} of Assignment-1 scores (\textit{$NC_1$}, \textit{$NC_2$}, \textit{$NC_3$}, with \textit{$NC_3$} denoting  the \textit{NC} of highest performing and \textit{$NC_1$} denoting the \textit{NC} of lowest performing students). Based on Assignment-1 score terciles, students were evenly distributed across each condition for reflection-1 (\textit{AI-hint-revision}: \textit{$NC_1$}=0.22, \textit{$NC_2$}=0.28, \textit{$NC_3$}=0.50; \textit{Hint-writing}: \textit{$NC_1$}=0.19, \textit{$NC_2$}=0.29, \textit{$NC_3$}=0.52 ). However, the distribution for reflection-2 was relatively uneven (\textit{AI-hint-revision}: \textit{$NC_1$}=0.17, \textit{$NC_2$}=0.33, \textit{$NC_3$}=0.50; \textit{Hint-writing}: \textit{$NC_1$}=0.25, \textit{$NC_2$}=0.25, \textit{$NC_3$}=0.50). This was because some students completed only one of the two reflections.


\subsection{Study Results}
\label{study_results}
We evaluated the GPT-4 hints based on the criteria described in Section \ref{measures}. The GPT-4 hint for reflection-1 was fully accurate, specific, well phrased, but had some extraneous information. Therefore, it was rated 6 for \textit{Utility}. In contrast, while the GPT-4 hint for reflection-2 was well phrased and specific, it was only partially accurate and consisted of extraneous information. Since this hint identified the primary mistake but required modification to remove the extraneous information, it was rated as 4 for \textit{Utility}. Importantly, unlike the extraneous information in reflection-1's GPT-4 hint, the extraneous information in reflection-2's GPT-4 hint was potentially distracting for students' learning. For comparison with the learnersourced hints, Table \ref{tab:desc} includes the applicable ratings for each GPT-4 hint.

We now discuss the study results based on the outcome variables described in Section \ref{measures}. Compared to the students in the \textit{Hint-writing} condition, students in the \textit{AI-hint-revision} condition wrote hints with higher mean values of \textit{Full Accuracy}, \textit{Specificity}, and \textit{Utility} in each reflection, and higher mean value of \textit{Phrasing} in reflection-2 (see Table \ref{tab:desc}). Regarding \textit{Extraneous Information}, students in the \textit{AI-hint-revision} condition had a higher mean in reflection-1, but a lower mean compared to the students in \textit{Hint-writing} condition in reflection-2.
Students in the \textit{AI-hint-revision} condition also spent more time writing hints in each reflection. Not all of these results were statistically significant. 

\begin{figure*}[t]
  \centering  \includegraphics[width=0.68\columnwidth]{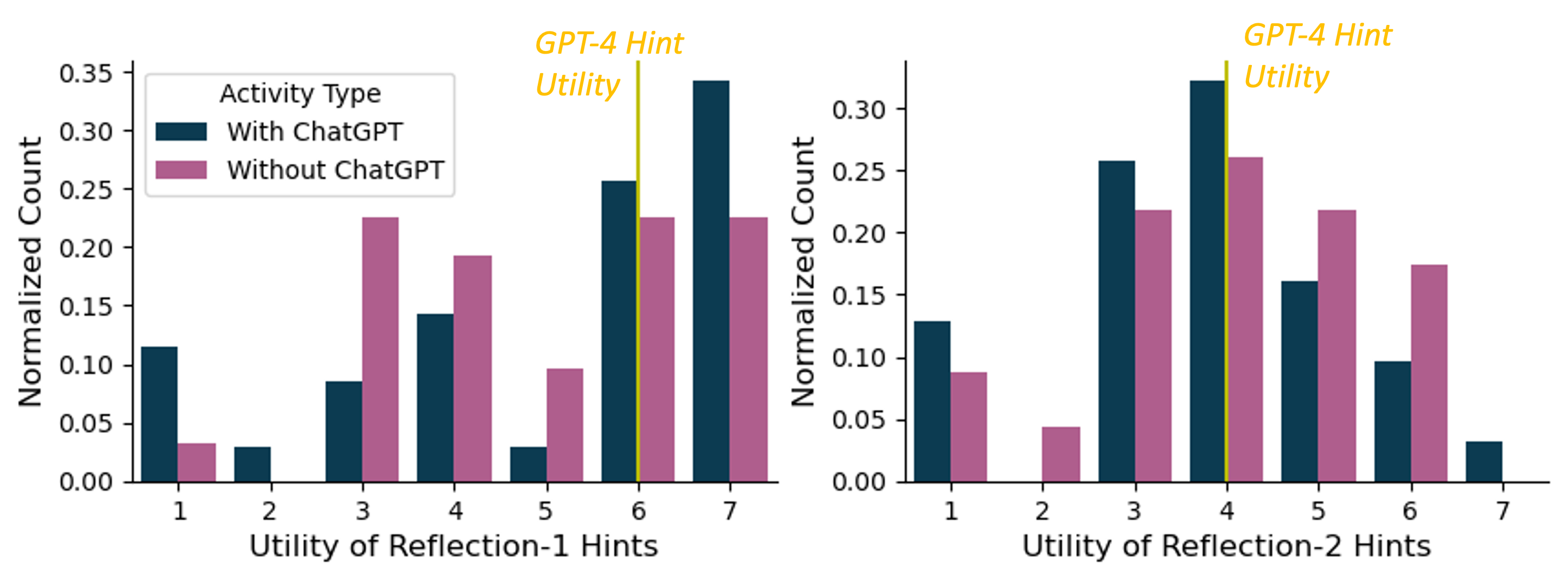}
  \vspace{-2mm}
  \caption{Normalized counts of students per each level of \textit{Utility} (with `7' representing most usable and `1' representing least usable hints) for each reflection. The yellow vertical lines indicate the \textit{Utility} of the GPT-4 hint for each reflection.} 
    \label{utility_plots}
    \vspace{-4mm}
\end{figure*}

We now report the results for those outcome variables which were significantly impacted by the factors \textit{condition}, \textit{course performance}, and/or the \textit{reflection number}\footnote{The results of all the statistical analyses are provided in detail \href{https://osf.io/vgrfk/?view_only=8579adb171a64c07bb6641be67dba202}{here}}. We use p<0.05 as the criterion for assessing statistical significance. 
The generalized linear mixed effects model analyses (using binomial family distribution with a logit link) for the outcome variables with binary values revealed the following: (i) The effect of reflection number on the log-odds of \textit{Full Accuracy} was statistically significant ($\hat{\beta}$ = -1.79, SE = 0.54, p < 0.01). Post-hoc pairwise comparisons with Tukey's adjustment confirmed that students wrote significantly more fully accurate hints in reflection-1 compared to reflection-2 for both the \textit{Hint-writing} (p=0.001) and \textit{AI-hint-revision} (p=0.001) conditions. (ii)  Students wrote significantly better phrased hints in reflection-1 compared to reflection-2 ($\hat{\beta}$ = -6.40, SE = 2.96, p < 0.05). Additionally, the model suggested that condition had a significant impact on the log-odds of phrasing ($\hat{\beta}$ = 8.09, SE = 3.39, p < 0.05) and that the interaction between condition and students' course performance was significant ($\hat{\beta}$ = -5.36, SE = 2.20, p < 0.05). However, post-hoc comparisons with Tukey's adjustment did not reveal a significant difference (at our threshold of p<0.05) between conditions for phrasing when evaluated at the mean value of course performance (p=0.1017). The Cumulative Link Mixed Model revealed that the hints written for reflection-1 had significantly higher \textit{Utility} than those written for reflection-2 ($\hat{\beta}$ = -1.85, SE = 0.40, p < 0.001). Further, students' course performance had a significant impact on hint utility ($\hat{\beta}$ = 0.58, SE = 0.24, p < 0.05). Fig. \ref{utility_plots} shows the normalized counts of students per each level of \textit{Utility} for each reflection. For comparison, the yellow vertical lines indicate the \textit{Utility} of the GPT-4 hint for each reflection.

\begin{figure*}[t]
  \centering
  \includegraphics[width=\columnwidth]{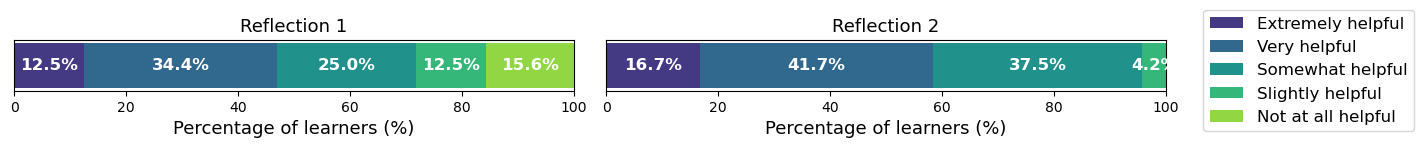}
  \caption{Students' responses regarding the extent to which the \textit{AI-hint-revision} variant of the exercise helped them think critically about the accuracy and appropriateness of GPT-4 responses}
    \label{LLM-critical}
\end{figure*}

In response to the question ``Would you be interested in doing this exercise in a future course?", 86.96\% students from the \textit{Hint-writing} condition and 71.43\% students from the \textit{AI-hint-revision} condition said ``Yes" after reflection-1, and 75\% students from the \textit{Hint-writing} condition and 91.3\% students from the \textit{AI-hint-revision} condition said ``Yes" after reflection-2. For the question, ``How helpful was this exercise in encouraging you to think critically about the accuracy and appropriateness of responses provided by GPT-4?", 71.9\% students in reflection-1 and 95.9\% students in reflection-2 found the exercise with GPT-4 hint at least somewhat helpful for thinking critically about LLM responses. The median ratings were 3 and 4 on a 5 point Likert scale for reflection-1 and reflection-2, respectively. Fig. \ref{LLM-critical} shows the distribution of students' responses to this question.
In response to the question ``Did you like doing this exercise with or without the GPT-4 hint?'', 54.9\% students indicated ``With GPT-4" and the remaining indicated ``Without GPT-4". 

Based on their open-ended responses regarding the reasons behind their choice, we found the following reasons in favor of engaging in the hint writing activity supported by a GPT-4 hint: (1) Students were curious about the capabilities of GPT-4 and felt that this exercise pushed them to think critically about its abilities and limitations (``\textit{This [exercise] is a great addition [to the course] because it takes into account the modern reality of students today and how they can best use the tools available to them to learn something new}"). (2) Students found the GPT-4 hint helpful for writing their own hint as it provided guidance with framing the hint and identifying the mistakes in the incorrect code (``\textit{I think having a certain direction to go in based on GPT-4 hint allows me to dive deeper into those concepts with more specificity and also primes me to do my own work rather than taking GPT-4 hints at face value.}"). In contrast, the students who indicated that they would prefer to write hints without the GPT-4 hint mentioned the following reasons: (1) Students did not trust GPT-4's responses and felt that the GPT-4 hint could be misleading (``\textit{I feel like GPT-4 sometimes give a hallucinated codes and answers}"). (2) Students felt that the GPT-4 hint biased the way they thought about writing their own hint (``\textit{GPT-4 pigeonholes me into a thought process that involves everything that GPT-4 has included in their hint, be it right or wrong}"). (3) Students felt that including the GPT-4 hint hinders learning and would have liked to write a hint without any external help (``\textit{The GPT-4 response takes away from the initial investigation I do to understand all mistakes in the code. I like doing the initial investigation by myself first and then maybe seeing a GPT-4 response.}"). (4) Students felt that the GPT-4 hint increased the task complexity (``\textit{ChatGPT makes me confused.}"). 

\section{Discussion and Future Work}
Regarding RQ1, on the difference between the quality of hints written by learners with versus without the GPT-4 scaffold, we found that the mean usability of hints written with the GPT-4 scaffold was higher than the hints written without it. However, this result was not statistically significant. This could be due to the current scaffold design or low statistical power given our chosen threshold of significance (p<0.05), which is a limitation of our study.
We also found that students who revised the GPT-4 hints wrote relatively better phrased hints compared to those who wrote it independently. This result was marginally significant and should be carefully considered in follow-up investigations. An intuitive finding,
Based on Fig. \ref{utility_plots} and Table \ref{tab:desc}, it is worth noting that most students in the \textit{AI-hint-revision} condition wrote hints with \textit{Utility} values that were close to the GPT-4 hint's \textit{Utility}. 

To interpret these results, we first need to understand the differences in the quality of the GPT-4 hints for each reflection. 
There were fewer mistakes in reflection-1’s incorrect solution compared to reflection-2. Moreover, the problem statement for this week’s programming assignment reminded students to label the axes and the plots appropriately, as this was the first assignment in the course. In contrast, the problem statement for reflection-2 did not remind students of these data visualization basics, which made it more difficult. It is likely that these differences in the problem statements (which were included as is in the GPT-4 prompts) and their difficulty levels contributed to the varying levels of usability of GPT-4 hints. Additionally, the extraneous information in the GPT-4 hint for reflection-2 could have been distracting for the learners (as mentioned in Section \ref{study_results}). This could explain why, in reflection-1 (where the GPT-4 hint was more usable) relatively more students were able to improve upon the quality of the GPT-4 hint compared to reflection-2 (where the GPT-4 hint was less usable), as shown in Table \ref{utility_plots}. This suggests that poor quality LLM hints can negatively impact the quality of learnersourced hints. This raises the need to explore techniques for selecting LLM hints that pass a given quality threshold, before showing to learners. Recent work \cite{phung2023generating} showing how educators can be given control over the quality of LLM-generated feedback before providing to students can be helpful in this regard. 


Regarding RQ2, on the relationship between students’ course performance and the quality of their hints, we found that higher performing students wrote more usable hints. This finding is corroborated by prior research highlighting that higher performing students write more comprehensive and accurate explanations \cite{chi1989self}. Future research can investigate further how student performance influences hint quality when written with different scaffold designs.

Regarding RQ3, on students' preference for writing hints with or without the GPT-4 scaffold, we did not find a clear preference for either design. On probing the reasons behind students' preferences, students' lack of trust in LLMs and willingness to engage in this activity independently, without getting support or getting biased, emerged as prominent reasons for preferring to write hints without the GPT-4 scaffold. This could be explained by the background of students in the data visualization course, who are all adult learners. 
The students enrolled in the graduate degree program that this course is part of typically demonstrate high levels of motivation for learning data science, as many of them are interested in switching career paths. This could explain their interest in engaging in this activity without any additional scaffolding. Their lack of trust in GPT-4 highlights the need for instructors to elaborate on the steps taken to mitigate any harmful effects of LLM responses, such as being transparent about the prompting strategies used.

Regarding RQ4 and in terms of student learning, the hint writing activities gave learners an opportunity to actively engage with LLMs and encouraged the majority of them to think critically about GPT-4's abilities. More research is needed to understand how to \textit{design} and \textit{present} such scaffolds to students based on their motivation levels and abilities, to concurrently help them learn deeply and write high quality hints. 

Taken together, these results suggest that LLM-generated hints can serve as useful scaffolds for students to write high quality hints, particularly when the LLM-generated hints are highly accurate. 
However, it is important to be careful about the design of such scaffolds. For instance, students can be given more agency in hint-writing by providing the AI-scaffold \textit{after} they have written a hint on their own to provide them with a `second opinion’. This design has been found useful for improving the quality of generated artifacts in prior research on learnersourcing \cite{choi2022algosolve} and can be especially useful for highly motivated learners. Perhaps for learners with a broader range of motivation levels, such as MOOC learners \cite{singh2021s}, the AI-scaffold could be provided on demand. 


\section{Limitations and Conclusion}
This study contributes to research on understanding the impacts of LLMs in education, and how we can optimize student learning as students interact with LLMs. A limitation of the study is that for the second reflection activity, students were not evenly distributed across experimental conditions based on their prior knowledge. Additionally, the number of students needs to be increased to enhance the statistical power of the study. Finally, we focused specifically on a single course at a single institution, although, having two reflections based on two different assignments helped address this limitation to some extent. Overall, as LLMs continue to become more powerful and accurate with time \cite{bubeck2023sparks}, this work has promising implications for using them to scaffold learnersourcing tasks like feedback generation while providing students with an opportunity to learn from interacting with them.
\vspace{-3.5mm}

\bibliographystyle{ACM-Reference-Format}
\bibliography{sample-base}




\end{document}